\documentclass[letterpaper,conference]{IEEEtran}

\usepackage{pgfplotstable}
\usepackage{algorithm,algorithmic}
\usepackage{amsmath}
\floatname{algorithm}{Procedure}
\usepackage{xcolor}
\usepackage[hyphens]{url}
\definecolor{textblue}{rgb}{.2,.2,.7}
\definecolor{textred}{rgb}{0.54,0,0}
\definecolor{textgreen}{rgb}{0,0.43,0}
\usepackage{tcolorbox}
\usepackage{numprint}
\usepackage{pgfplots}
\pgfplotsset{compat=1.8}
\usepgfplotslibrary{groupplots}
\usepackage{relsize}
\usepackage{listings}
\usepackage{xspace}
\usepackage{amssymb}
\usepackage{pgfplotstable}
\usepackage{csvsimple}
\usepackage{verbatim}
\usepackage{array,booktabs}
\usepackage{adjustbox}
\usepackage{listings}
\usepackage{mathtools}
\usepackage{fancyvrb}
\usepackage{booktabs} 
\usepackage{siunitx} 
\usepackage{pgfplotstable} 
\usepackage{hyperref}
\usepackage{underscore}
\hypersetup{
    urlcolor=blue,
}
 
\usepackage{amssymb}

\def\ojoin{\setbox0=\hbox{$\bowtie$}%
  \rule[-.02ex]{.25em}{.4pt}\llap{\rule[\ht0]{.25em}{.4pt}}}

\def\rightouterjoin{\mathbin{\bowtie\mkern-5.8mu\ojoin}}

\newcommand{\cbox}[1]{\fcolorbox{red}{white}{\textcolor{red} {#1}}}

\definecolor{pblue}{rgb}{0.13,0.13,1}
\definecolor{pgreen}{rgb}{0,0.5,0}
\definecolor{pred}{rgb}{0.9,0,0}
\definecolor{pgrey}{rgb}{0.46,0.45,0.48}

\usepackage{tabularx,colortbl}
\usepackage{tikz}
\usetikzlibrary{patterns}

\newcommand{\code}[1]{\textit{#1}}
\begin{document}

\title{Understanding the Quality of Container Security Vulnerability Detection Tools}
\author{
 \IEEEauthorblockN{Omar Javed\IEEEauthorrefmark{1} and Salman Toor\IEEEauthorrefmark{2}, }
\IEEEauthorblockA{\IEEEauthorrefmark{1}Universit\`a della svizzera italiana, Lugano, Switzerland
 \\\{omar.javed\}@usi.com}
 \IEEEauthorblockA{\IEEEauthorrefmark{2}Uppsala University, Uppsala, Sweden
  \\\{salman.toor\}@it.uu.se}
  }

\maketitle
\thispagestyle{plain}
\pagestyle{plain}

\begin{abstract}

Virtualization enables information and communications technology industry to better manage computing resources. In this regard, improvements in virtualization approaches together with the need for consistent runtime environment, lower overhead and smaller package size has led to the growing adoption of containers. This is a technology, which packages an application, its dependencies and Operating System (OS) to run as an isolated unit. However, the pressing concern with the use of containers is its susceptibility to security attacks. Consequently, a number of container scanning tools are available for detecting container security vulnerabilities. Therefore, in this study, we investigate the quality of existing container scanning tools by proposing two metrics that reflects coverage and accuracy. We analyze \numprint{59} popular public container images for Java applications hosted on DockerHub using different container scanning tools (such as Clair, Anchore, and Microscanner). Our findings show that existing container scanning approach does not detect application package vulnerabilities. Furthermore, existing tools do not have high accuracy, since $\approx$34\% vulnerabilities are being missed by the best performing tool. Finally, we also demonstrate quality of Docker images for Java applications hosted on DockerHub by assessing complete vulnerability landscape  i.e., number of vulnerabilities detected in images.

 \end{abstract}

\section{Introduction}

Cloud based infrastructure alleviates the challenge of managing and maintaining application services across large distributed computing environments~\cite{Chhabra2015,montecchi}. However, the need for faster deployment, better performance and continuous delivery of application services has led to the introduction of containers~\cite{Martin2018}. 

Container is a virtualization approach that sits on top of a physical machine and shares its host OS kernel and services~\cite{Bernstein14}. The benefit of using containers over traditional virtualization approaches will lead to its growing adoption in the industry by 40\% in the year 2020~\cite{Applicat91}. Moreover, it is also expected that 47\% of information technology service providers are planning to deploy containers in their environment~\cite{httpsdia29}. 

One of the leading container technology is Docker that has more than 6 billion downloads\footnote{This figure was reported in 2016 --- \url{https://blog.docker.com/2016/10/introducing-infrakit-an-open-source-toolkit-for-declarative-infrastructure/}} and over million images on DockerHub\footnote{\url{https://www.docker.com/products/docker-hub}}. However, the popularity of Docker containers make them susceptible to security threats~\cite{Martin2018,duarte,Rui2017}. For example, in July 2017 it is reported that an attacker hosted several malicious Docker (container) images on DockerHub. Before these images were taken down, they were downloaded more than 5 million times, which resulted in 545 Monero digital coins being mined (approximately \$900,000)~\cite{Backdoor16}. 

To identify security issues, Docker Inc. has a scanning service~\cite{DockerSe92} which was formerly known as “ProjectNautilus”~\cite{dockerscan}. The service provides  automated monitoring, validation and detection of vulnerabilities for images hosted on DockerHub.  However, the service is currently not available for all Docker images (i.e., it does not scan community images). 
Hence, making community based Docker images crucial to investigate. 

Furthermore, a number of container scanning tools are available (e.g., Clair~\cite{clair}, Anchore~\cite{Anchore}, and Microscanner~\cite{microscan}) which can analyze official as well as community images hosted on DockerHub. The approach employed by these tools is that they collect package information (e.g., base OS\footnote{Base OS refers to an image that contains an operating system packaged in the Docker image.} packages) and compares it against a vulnerability database. To demonstrate vulnerability issues in both official and community images on DockerHub, Rui et al. analyzed OS packages~\cite{Rui2017}. However, the study did not explore the vulnerabilities detected in the application and its dependencies packaged in the container. 

Since there is a prevalence of different container scanning tools, it is important to assess the quality of these tools to better understand their strengths and weaknesses. Such kind of assessments are important to improve the approach being employed by detection tools. A high quality tool is important for addressing security issues because it can point to the component where the vulnerability exists, which can be fixed in a timely manner~\cite{BenOthmane2017}.

Therefore, in this study, we investigate the effectiveness of existing container scanning tools by proposing two metrics, which reflects tool's coverage and accuracy.  To the best of our knowledge, this is the first research study assessing container scanning tools on large number of real-world Docker images of Java-based container applications. We select Java applications because  being the most popular programming language~\cite{tiobeindex}, it is also susceptible to security attacks. For example, a new vulnerability related to remote code execution (CVE-2018-11776) was found in Apache Struts 2~\cite{CVE2018110}. This has affected many users and developers. It has been estimated that almost 65\% of Fortune 100 companies use Apache Structs~\cite{jaxenter}. This makes Java-based container applications important to study. 

Based on this premise, we formulate three research questions:

 \begin{itemize}
\item[] \textit{\textbf{RQ1: } Is existing container vulnerability scanning approach effective in finding nonOS vulnerabilities?} 
\item[] \textit{\textbf{RQ2: } What is the accuracy of vulnerability detection of existing container scanning tools? }
\item[] \textit{\textbf{RQ3: }What is the complete vulnerability landscape i.e., vulnerabilities detected in both OS and nonOS packages of community images for Java-based container applications on DockerHub?}
	
\end{itemize}

To answer these research questions, we analyze 59 popular Docker images of Java-based container applications hosted on DockerHub. This study is important and timely because it indicates shortcomings in existing container scanning tools, which we hope researchers and practitioners will improve. 
In this regard, our study makes the following key contributions:

\begin{itemize}

\item We present a large-scale evaluation of container vulnerability scanning tools by analyzing 59 popular public Docker images. 

\item We propose two metrics for assessing tool's quality which are coverage and accuracy of the container scanning tools.

\item We find that the application packaged in the container images are missed by container scanning approach. Thus, making detection coverage of the tools questionable.

 \item We demonstrate a high number of vulnerabilities being missed by the tools. Hence this affects the accuracy of the tool.
\item We provide a set of recommendations that can guide practitioners and researchers to develop and propose effective vulnerability detection approach for container scanning tools.

\end{itemize}

The rest of the paper is structured as follows:
Section~\ref{related work} discusses background and related work for this study.
The rationale for selecting the tools is presented in section~\ref{selection}, and section~\ref{metric} describes our proposed metrics for assessing tool quality.
Section~\ref{methodology} presents our approach for selecting and analyzing Docker images.
We describe our evaluation and findings in section~\ref{evaluation}.
Section~\ref{discussion} discusses our findings.
In section~\ref{threadvalid} we discuss threats to validity of our study .
We provide recommendations for future studies in section~\ref{recommendation} and we conclude in section~\ref{conclusion}.

\textbf{Dataset	} An important aspect of this study is the complete analysis of the vulnerability for its truthfulness. We provide complete information of detected vulnerabilities, along with list of all analyzed Docker images with its project information~\cite{dataset}.

\section{Background}
\label{related work}
In this section, we first explain a container scanning process, followed by discussing existing container vulnerability detection tools and services. Furthermore, we will also discuss  studies that investigate vulnerabilities in Docker images. Finally, we provide the rationale for selecting the different container scanning tools in our evaluation.

\begin{figure}[ht]
	\centering
	\includegraphics[scale=0.6]{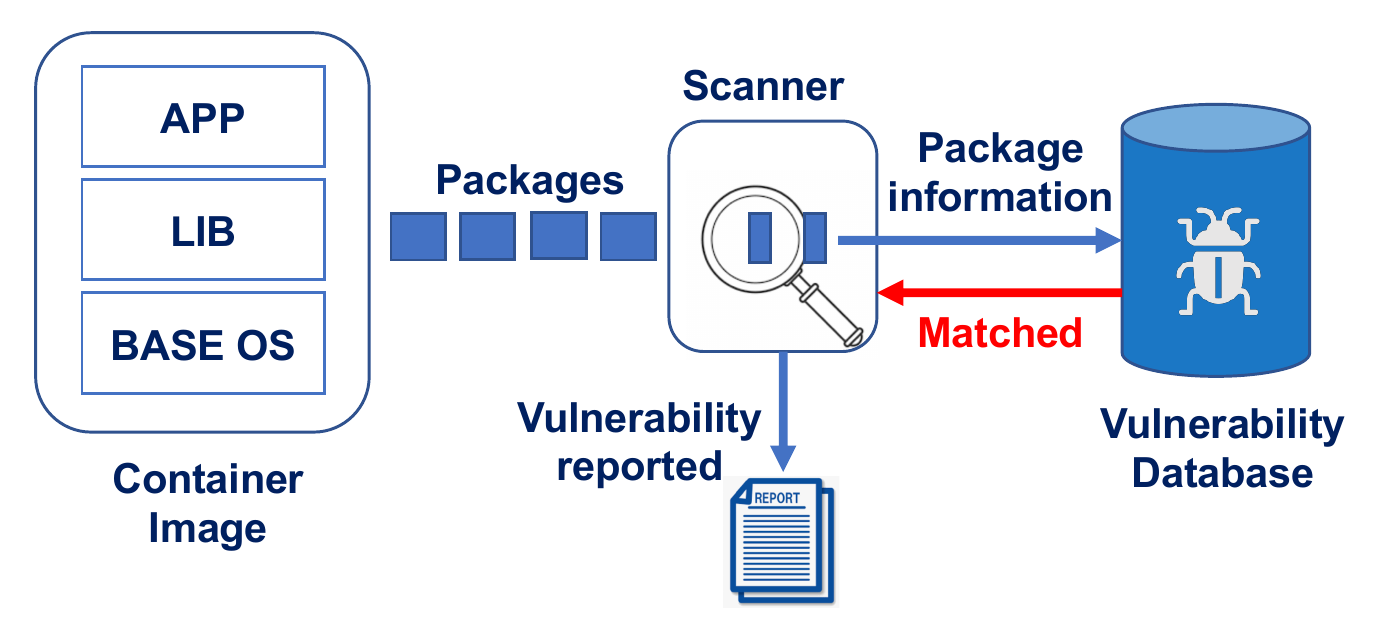}
	\caption{A typical container scanning approach for package vulnerability detection} 
	\label{fig:approach}
\end{figure}

\subsection{Container scanning process}

In a typical container scanning process, a number of packages in a container image are scanned by the tool, which analyzes the package name and its version number. This information is then compared against a vulnerability database, which contains a list of entries from publicly known security attacks (or exposures) also known as Common Exposures and Vulnerabilities (CVE). If the analyzed package and its version matches the entry in the database, this is reported as a vulnerability. The scanning process is shown in figure~\ref{fig:approach}.

Based on this idea, several different types of container vulnerability assessment tools and services have been provided, such as official Docker scanner, GitHub security alerts and  enterprise service (e.g., RedHat Containers).  Docker Inc. has a scanning service~\cite{DockerSe92}, which provides monitoring, validation and identification of vulnerabilities for official container images hosted on DockerHub. However, the service is currently not available for community-based images.  Similarly, GitHub also provides a service for alerting about vulnerabilities in the dependencies of a project~\cite{gitsecalarm}. However, the service is relatively young, and is expanding to support multiple languages. OpenSCAP~\cite{Security31} provides multiple utilities for determining the compliance of containers.  One of which is to scan Docker images for detecting vulnerability against CVE  database. However, the functionalities are only limited to Red Hat Docker Containers. Similarly, IBM's Vulnerability Advisor~\cite{Vulnerab81} also provides basic policies for security. However, it is only capable of analyzing images hosted on Bluemix cloud by monitoring and inspecting images against a vulnerability database.

\begin{table*}[ht]
\catcode`"=9			
\caption{List of Docker scanning tools (sorted by name of the tool)}
\centering
\begin{filecontents*}{./table/data/toolcomp.csv}
name,language,functionality
Anchore~\cite{Anchore},Python,Identify OS and nonOS vulnerabilities in Docker images
AppArmor~\cite{AppArmor83},C,Prevent access to filesystem and network
Cilium~\cite{cilium92},Go,HTTP-layer and network layer security detection
Calico~\cite{calico},Ruby,Virtual networking security detection
Clair~\cite{clair},Go,Identify OS vulnerabilities in Docker images
Dagda~\cite{eliasgra},Python,{Identify package vulnerabilities, trojan, viruses and malware in Docker images}
Notary~\cite{notary},Go,Verify the integrity and origin of the content
Microscanner~\cite{microscan},Dockerfile,Identify OS vulnerabilities in Docker images
Open Policy Agent~\cite{openpolicy},Go,Security policy enforcement
\end{filecontents*}
\pgfplotstabletypeset[
  col sep=comma,
  columns/name/.style={
  		string type,
  		column name=\textbf{Name}, 
		column type={l}
  },
  columns/language/.style={
  	string type,
	column name=\textbf{Language}, 
	column type={c}
	},	
  columns/functionality/.style={	
    	string type,
	column name=\textbf{Functionality},
    	column type={l}
  },
  every head row/.style={
  					before row={		
								\hline
							  },
					after row=\hline
				     },
  every last row/.style={after row=\hline},
  my row iterator/.style={every row no #1/.style={
    before row={
      \rowcolor{black!30}
      }
    }
  },
  my row iterator/.list={0,4,5,7}
  ]{./table/data/toolcomp.csv}

\label{table:toolcomp}
\end{table*}

\subsection{Docker image analysis}
Previous studies have conducted Docker image vulnerability analysis. Martin et al.~\cite{Martin2018} provided an assessment based on literature survey regarding the exploitation and mitigation of Docker image from an external attack. However, the study did not assess real-world Docker images hosted on DockerHub to identify security vulnerabilities that can potentially be exploited by an attacker. 

Rui et al.~\cite{Rui2017} conducted a study on security vulnerabilities in both official and community images on DockerHub. However, the study analyzed vulnerabilities that are related to only OS packages, and did not study the vulnerabilities that are present in nonOS packages. Furthermore, the evaluation and the analysis was conducted on one container scanning tool (i.e., Clair).
We, on the contrary, analyze vulnerabilities detected in both OS and nonOS packages. Furthermore, we also investigate the effectiveness of the existing container scanning approach.

\section{Tool selection}
\label{selection}
Table~\ref{table:toolcomp} lists different types of Docker container scanning tools. We explored a number of different container scanning tools from different sources~\cite{10top,29Docker,Containerscan}, and identified around 30 container scanning tools. Commercial tools are excluded because complete features of the tools are not available in a trail version. Therefore, it will not provide an accurate comparison of the tool. Furthermore, we also do not report tools which are integrated in another scanning tool. For example, Dagda uses Sysdig Falco~\cite{FalcoSys23} for monitoring anomalous activities in Docker containers, we only report Dagda.

Furthermore, we assess only those container scanning tools, which detects package vulnerabilities. This is because such vulnerabilities are of high concern when deploying containers~\cite{Rui2017}. In table~\ref{table:toolcomp}, we highlight (in dark grey color) tools whose functionality is to detect vulnerabilities in OS and nonOS packages. From the table, we can see that out of 9 tools, only two tools identify vulnerabilities in nonOS packages (i.e. Anchore \& Dagda), which shows that there is a lack of container scanning tools that detect vulnerabilities in application and its dependencies packaged in the container. Furthermore, we find that Dagda~\cite{eliasgra} suffers from  inaccuracy of vulnerability detection as mentioned on Dagda project's issue page~\cite{Improvet85}, which leads to the exclusion of this tool from our analysis.
Based on these findings, we select Clair, Anchore, and Microscanner for our analysis. 

Clair is used by a public registry such as Quay.io for analyzing images to detect known vulnerabilities~\cite{QuaySecu15}. In our study, we use clair-scanner which eases the setup for scanning images. Clair analyzes the image by comparing detected package against vulnerability data  sources such as Ubuntu trackers, Debian trackers, RedHat security data, Oracle Linux security data, Amazon Linux security advisories, Alpine security database and NIST National Vulnerability Database (NVD). 

On the other hand Anchore has an engine, which like Clair, also uses security vulnerabilities from different sources. It detects OS package vulnerabilities from specific Linux distribution feeds, and NVD.

Microscanner has different versions available e.g., free and paid versions. The features available in free version are adequate for our evaluation~\cite{microscan} (e.g., package vulnerability detection). Furthermore, Microscanner also has a wrapper\footnote{https://github.com/lukebond/microscanner-wrapper}, which eases the scanning of images.

\section{Metric selection for assessing tool quality}
\label{metric}
Our study proposes three different metrics to assess the quality of  container scanning tools. We define and explain the reason for selecting each of the metrics as follows:

\textbf{Detection coverage:} The lack of container scanning approach analyzing nonOS package of the container is the main reason for selecting this metric. It is important to assess package (or a section of a container image) that is missed by the existing container vulnerability scanning approach. There are three different categories of packages that are contained in a container --- Application, Dependencies (or library), and OS packages. We investigate whether existing container scanning approach is feasible to detect all three different category of packages. Hence, the metric indicates tool \textit{coverage}.

\textbf{Detection Hit Ratio (DHR):} This metric demonstrates the tool's effectiveness in terms of vulnerability detection, i.e., the number of vulnerabilities successfully detected by a tool from a given set of vulnerabilities.  The higher the detection, the better is its effectiveness.  
Therefore, this metric indicates tool \textit{accuracy}. Furthermore, the important aspect of computing detection hit ratio is the number of detection misses, therefore, we explain our procedure for finding detection miss for each tool in section~\ref{miss}
  We formulated Detection Hit Ratio (DHR) to measure the tool's accuracy. DHR provides an indication of how many vulnerabilities are detected from a given set of vulnerabilities. We compute DHR by using the following formula:

\[
  \text{DHR} = \frac{\text{Detection Hit}} {(\text{Detection Hit + Detection Miss})}
\]

where
\begin{description}
\item[---]  Detection Hit is the number of vulnerabilities detected.
\item[---]  Detection Miss is the number of vulnerabilities missed. 
\end{description}

\section{Methodology}\label{methodology}

\begin{figure}[ht]
	\centering
	\includegraphics[scale=0.48]{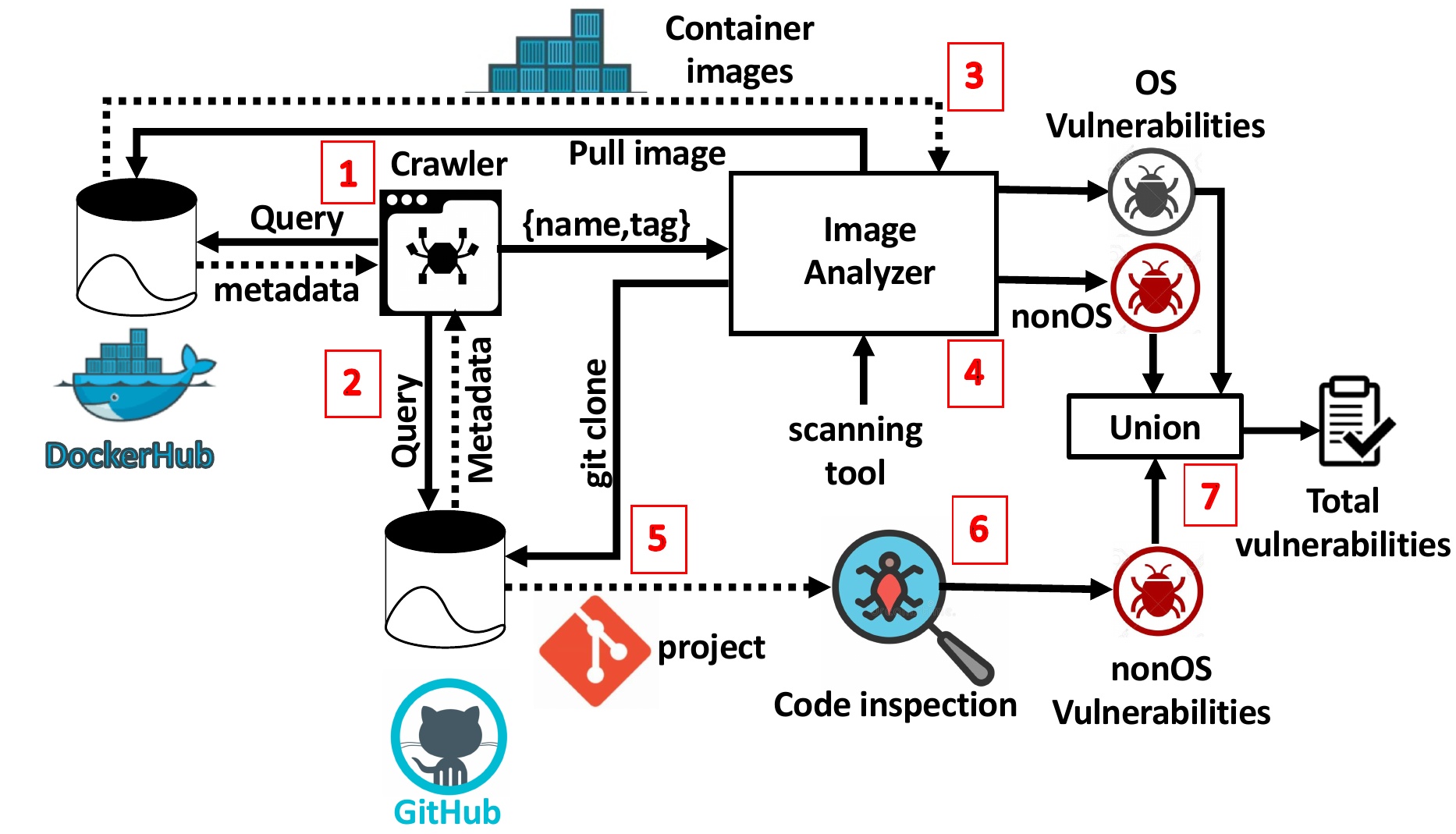}
	\caption{Methodology for collecting and analyzing Docker images} 
	\label{fig:methodology}
\end{figure}

We present a methodology to automatically find a set of popular Docker images.  
Our methodology is based on collecting images from DockerHub and their corresponding projects\footnote{We refer to projects as GitHub repository of the Docker image.} on GitHub. The images are analyzed by container scanning tools, and image's corresponding project code is used for code inspection by using SpotBugs\footnote{https://find-sec-bugs.github.io/}) for finding security bugs in the application code. Fig.~\ref{fig:methodology} illustrates our methodology, which we explain in subsequent sub-sections.

\subsection{Finding Docker image for Java-based container applications}
\begin{figure}[ht]
	\centering
	\includegraphics[scale=0.56]{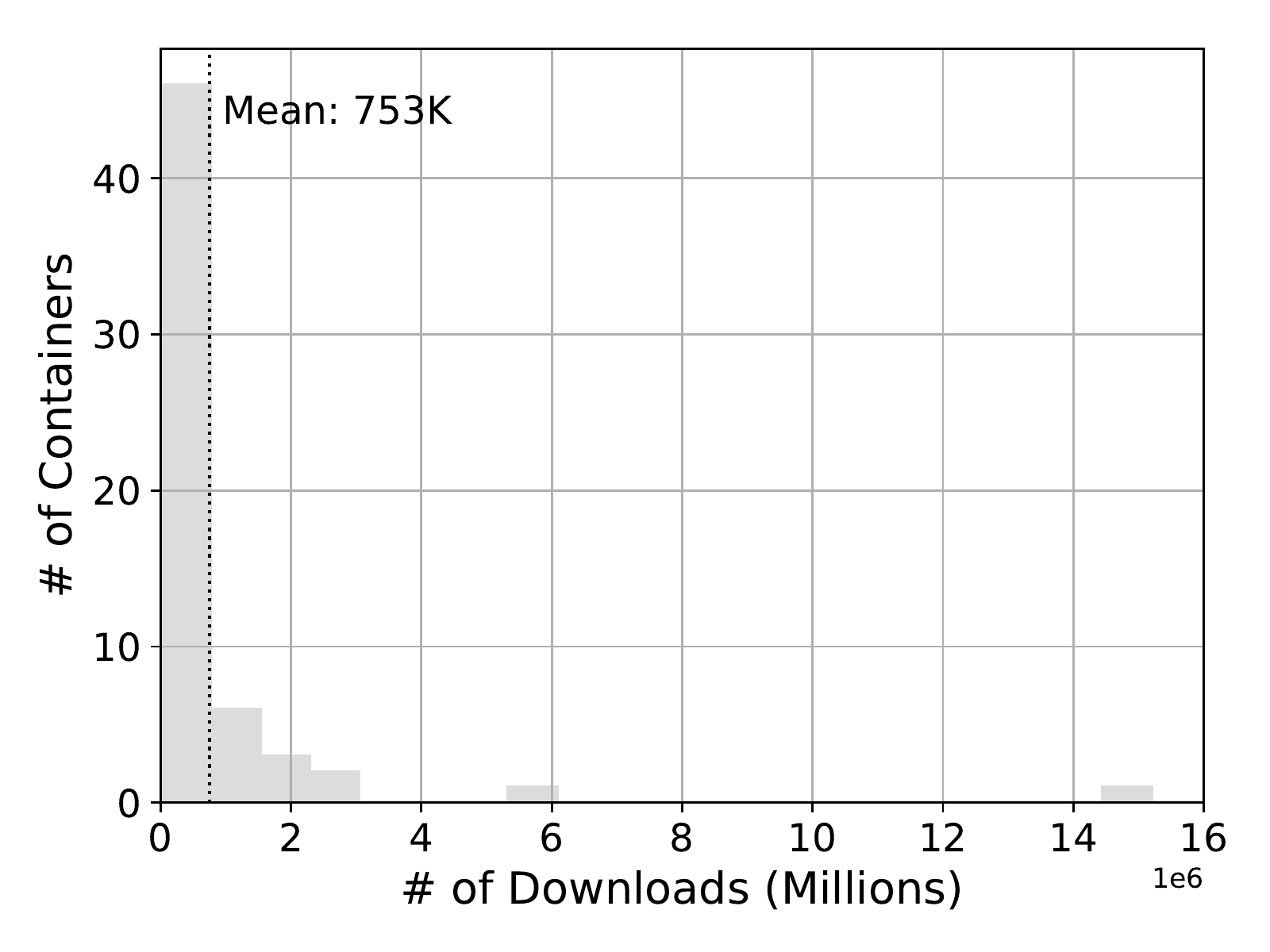}
	\caption{Histogram representing number of downloads of Docker images. } 
	\label{fig:histogram}
\end{figure}

We first query DockerHub API to find a set of popular public (i.e., community-based) Docker images\footnote{Search API of DockerHub returns official images, therefore, in order to find public images, we specify \code{``community''} type in the \code{source} parameter of the API.} (see \cbox{1} in figure~\ref{fig:methodology}). In order to find  the corresponding image source, we query build information\footnote{\url{https://hub.docker.com/v2/repositories/${name}/autobuild/}} of the image, which provides the corresponding project's GitHub link. We then query GitHub API (see \cbox{2}  in figure~\ref{fig:methodology}) to select only Java projects. This provides a set of Docker images for Java-based container applications. We store the set of images in the form of image name and its corresponding tag. We call this data collection phase as ``crawler", which we depict in figure~\ref{fig:methodology}. The popularity of crawled Docker images for Java-based container applications is shown in figure~\ref{fig:histogram}. This popularity is described in terms of number of downloads (as shown in the x-axis). The mean download of the images are in the order of \numprint{750000}. This shows that images in our study are being used by many users or developers.  Furthermore, around 10 Docker images are in the order of 1 to 15 million downloads.

\subsection{Analyzing Docker images for detecting vulnerabilities}
To detect vulnerabilities, we analyze the set of Docker images using container scanning tools such as Clair, Anchore and Microscanner. We refer to this phase as ``Image analyzer", which downloads (i.e., pull) the image from DockerHub (see \cbox{3} in figure~\ref{fig:methodology}).  To detect OS and nonOS package vulnerabilities, these images are fed to a container scanning tool (as shown by \cbox{4} in figure~\ref{fig:methodology}). A vulnerability report is generated per image, which is reported in Java script object notation form. Therefore, these reports can be saved on the disk or easily exported to document-oriented database such as MongoDB.  

\subsection{Code level inspection  for vulnerability detection}

We use security plugin of SpotBugs (a static code analyzer) to detect vulnerabilities in the application code of the container. This step is important for assessing ``detection coverage'' of the container scanning tool. Therefore, we first download the project repository from GitHub (see \cbox{5} in figure~\ref{fig:methodology}) and then compile its code. This is followed by the analysis of SpotBug's security plugin on the project's compiled code (see \cbox{6} in figure~\ref{fig:methodology}).

Finally the result of the two analyses i.e., one with the container scanning tool  and the other one with SpotBugs are combined to produce the total number of vulnerabilities detected in both OS and nonOS packages (see \cbox{7} in figure~\ref{fig:methodology}). This would assess ``complete vulnerability landscape'' of the tools.

\section{Evaluation}\label{evaluation}

\subsection{Evaluation Setup}\label{setup}

The analysis of this study has been conducted on Ubuntu Linux 16.04.4 LTS operating system. Furthermore, the hardware configuration of the machine is 8x 2.2 GHz CPUs, 16 GB of RAM. Based on our exploratory analysis  in section~\ref{selection}, we assess three open source container scanning tools (i.e., Clair, Anchore, and Microscanner). For code level inspection, we use a security plugin of SpotBugs version 1.9.0 released on March 27th, 2019.

\subsection{Identifying vulnerabilities in  nonOS packages with existing container scanning approach}\label{nonOSanalysis}

Docker containers provide a way to package an application, which can run on a cluster, cloud, or even on a standalone machine. The application is packaged into a container that uses a number of Open-Source Software (OSS) libraries. According to a report, 80\% to 90\% of applications use OSS library's components~\cite{TheState20}. Therefore, we scan Docker images to detect nonOS package vulnerabilities (i.e., OSS libraries used by the container-based application). We find that only Anchore can analyze nonOS packages.\footnote{To detect nonOS packages used by the container-based application, we selected ``non-os'' option during image scan. This will exclude other type of vulnerabilities such as vulnerabilities from OS.}. Clair and Microscanner does not have nonOS package vulnerability feeds. 

We analyze the latest version of Docker images. The version is  identified by leveraging the DockerHub convention of using ``imagename:tag''. Here ``imagename'' is the name of the image (e.g., apache/nutch) and  ``tag'' is either the \textit{latest} or any other name specified by the developer of the image (e.g., a version number).

While crawling of DockerHub images, we search for the image's most recent tag name\footnote{\url{https://registry.hub.docker.com/v2/repositories/${imagename}/tags/}}. The identification of tag is necessary because Anchore fails to analyze an image if correct tag value is not provided. Furthermore, we find that not all images use ``:latest'' as a tag to indicate its most recent version. Out of 59 images, we find that 37 images use ``:latest'' tag for providing its latest image version. Therefore, based on the correct tag value, we analyze the most recent version of 59 Docker images to detect vulnerabilities in nonOS packages.

\begin{table}[ht]
\caption{Severity level ranked by number of detected vulnerabilities}
\centering
\begin{filecontents*}{./table/data/app-severity-frequency.csv}
severity,frequency
Medium,10080
High,3316
Low,1137
\end{filecontents*}
\pgfplotstabletypeset[
  col sep=comma,
  columns/severity/.style={
  		string type,
  		column name=\textbf{Severity level}, 
		column type={l}
  },
  columns/frequency/.style={
  	string type,
	column name=\textbf{\# Detected Vulnerabilities}, 
	column type={c}
	},	
 every head row/.style={
  					before row={		
								\toprule
							  },
					after row=\hline
				     },
  every last row/.style={after row=\bottomrule},
  ]{./table/data/app-severity-frequency.csv}

\label{table:sevtype}
\end{table}

Table~\ref{table:sevtype} shows different severity levels for the vulnerabilities detected. We find in total \numprint{14533}   
vulnerabilities in 800 packages. Out of \numprint{14533}, \numprint{1137} are low severity level vulnerabilities, meaning that they not pose any significant threat and are not exploitable  by an attacker. Therefore, after filtering low severity level vulnerabilities, there are still \numprint{13396} vulnerabilities left in \numprint{795} packages. The presence of these vulnerabilities make these Docker images susceptible to potential threat.

The process of confirming bugs in a large-scale study is quite challenging as it requires an understanding of different domain application~\cite{Legunsen16}. Furthermore, static analysis can generate false warning for vulnerability detection~\cite{thome}. To ease the process,  we  inspect  vulnerabilities by checking National Vulnerability Database (NVD) for each reported CVE identifier (such as https://nvd.nist.gov/vuln/detail/{CVE identifer})~\footnote{NVD maintains information about each reported vulnerability.}. This allows us to understand whether the vulnerability is fixed or not.  Hence, we only report nonOS package vulnerabilities whose fix are available i.e., problem has been acknowledged by the developer and whose solution has been provided  in the form of a patch or package update. Therefore, out of \numprint{13396}, we find  \numprint{1039} vulnerabilities whose fixes are not available. These vulnerabilities were present in 55 (out of 59) docker images, which are reported in table~\ref{table:mainvultable}.  
 
\begin{table*}[ht]
\caption{List of Vulnerabilities reported by Anchore in nonOS packages --- application and dependencies in 55 (out of 59) Docker images.}
\centering
\begin{filecontents*}{./table/data/main-table.csv}
ID,DockerImage,VulApp,VulDep,DetectedVul,HighVulInPackage,MostVulPackage
P1,88250/solo:v3.0.0,0,4,128,120,mysql-8.0.12
P2,activiti/activiti-cloud-gateway:7.1.18,0,3,8,6,jackson-databind-2.9.9
P3,activiti/activiti-cloud-modeling:7.1.357,0,3,8,6,jackson-databind-2.9.9
P4,adrien1251/spi:latest,0,30,898,494,tomcat-8.5.11
P5,allegro/hermes-management:hermes-0.12.1,0,13,121,69,tomcat-8.5.5
P6,annoying34/annoying34-backend:latest,0,12,783,540,mysql-5.1.13
P7,apache/bookkeeper:4.9.1,0,3,13,11,jackson-databind-2.9.7
P8,apache/nutch:latest,0,38,554,63,jackson-databind-2.9.0
P9,appuio/example-spring-boot:renovate_configure,0,19,970,524,mysql-5.1.38
P10,bo0tzz/complimentbot:latest,0,13,32,10,maven-3.0
P11,bo0tzz/potatosbot:latest,0,13,34,10,maven-3.0
P12,bo0tzz/remindmebot:latest,0,14,35,10,maven-3.0
P13,bo0tzz/topkekbot:latest,0,14,36,10,maven-3.0
P14,bptlab/argos-backend:2.1.1,0,3,3,1,maven-3.1
P15,broadinstitute/gatk:latest,0,13,105,60,hadoop-2.8.2
P16,broadinstitute/picard:2.20.8,0,13,63,41,gradle-5.6
P17,buremba/rakam:latest,0,53,742,522,mysql-5.1.44
P18,coco/methode-article-mapper:latest,0,6,22,9,jackson-2.3.3
P19,consol/sakuli-ubuntu-xfce:dev,0,8,287,261,mysql-5.1.25
P20,dfbnc/dfbnc:latest,0,11,65,48,gradle-4.4.1
P21,emcmongoose/nagaina:3.6.11,0,1,21,21,jackson-databind-2.9.1
P22,exemplator/exemplator:latest,0,6,37,15,jackson-databind-2.6.2
P23,garystafford/voter-service:0.2.110,0,6,106,69,tomcat-8.5.6
P24,gazgeek/springboot-helloworld:latest,0,25,536,324,tomcat-8.0.20
P25,goeuro/opentripplanner:1.1.0,0,12,36,11,mercurial-3.1.2
P26,hashmapinc/drillflow:latest,0,3,24,12,tomcat-9.0.12
P27,hawkbit/hawkbit-update-server:0.3.0M4,0,1,7,7,jackson-databind-2.9.8
P28,iotaledger/iri:dev,0,2,3,2,jackson-databind-2.9.9.2
P29,ismisepaul/securityshepherd:dev,0,2,262,261,mysql-5.1.24
P30,jaegertracing/spark-dependencies:latest,0,8,177,112,hadoop-2.6.5
P31,jamesdbloom/mockserver:mockserver-snapshot,0,3,9,6,jackson-databind-2.9.9
P32,jmxtrans/jmxtrans:latest,0,5,26,22,jackson-databind-2.8.9
P33,jrrdev/cve-2017-5638:latest,0,12,206,100,tomcat-8.0.33
P34,jwhostetler/guacamole:latest,0,9,739,324,tomcat-8.0.20
P35,kennedyoliveira/hystrix-dashboard:dev,0,17,93,66,jackson-databind-2.7.4
P36,kromit/rkhelper:latest,0,25,94,44,jackson-databind-2.7.2
P37,maite/application:latest,0,30,900,494,tomcat-8.5.11
P38,mariobehling/loklak:development,0,5,16,8,jackson-databind-2.2.3
P39,meirwa/spring-boot-tomcat-mysql-app:latest,0,31,1085,528,mysql-5.1.31
P40,nmdpbioinformatics/phycus:latest,0,19,287,176,mysql-8.0.13
P41,onosproject/onos:latest,0,7,31,18,jackson-databind-2.9.5
P42,overture/score:edge,0,5,38,23,jackson-databind-2.8.3
P43,owasp/zap2docker-stable:latest,0,5,43,21,jackson-databind-2.9.8
P44,pelleria/hunter:latest,0,30,898,494,tomcat-8.5.11
P45,reportportal/service-api:5.0.0-BETA-12,0,8,33,12,tomcat-9.0.12
P46,schibstedspain/turbine-hystrix-dashboard:latest,0,11,174,108,tomcat-8.0.23
P47,shadovokun/obarbecue:latest,0,30,898,494,tomcat-8.5.11
P48,sonamsamdupkhangsar/springboot-docker:latest,0,13,169,95,tomcat-8.5.11
P49,srevereault/javalove:latest,0,1,11,11,mercurial-3.1.2
P50,strm/tasker:latest,0,4,34,22,jackson-databind-2.8.9
P51,surfsara/evidence-newsreader-k8s-rite:latest,0,56,244,81,batik-1.7
P52,tremolosecurity/kubernetes-artifact-deployment:latest,0,1,7,7,jackson-databind-2.9.8
P53,vidanyu/ache:latest,0,15,61,18,jackson-databind-2.9.5
P54,yegor256/rultor:1.68.2,0,47,139,30,batik-1.8
P55,yegor256/threecopies:latest,0,6,6,1,qs-6.2.1
\end{filecontents*}
\pgfplotstabletypeset[
  col sep=comma,
  font=\scriptsize,
  columns/ID/.style={
  		string type,
  		column name=\textbf{}, 
		column type={l}
  },
  columns/DockerImage/.style={
  		string type,
  		column name=DockerImage, 
		column type={l}
  },
  columns/VulApp/.style={
  		string type,
  		column name= Application,
		column type={r}
  },
  columns/VulDep/.style={
  	string type,
	column name= Dependencies , 
	column type={r}
	},	
  columns/DetectedVul/.style={	
    	string type,
	column name=Total Vulnerability,
    	column type={r}
  },
  columns/HighVulInPackage/.style={	
    	string type,
	column name= Highest,
    	column type={r}
  },
  columns/MostVulPackage/.style={	
    	string type,
	column name=Most Vulnerable Package,
    	column type={r}
  },
  every head row/.style={
  					before row={		
								\toprule
							  },
					after row={
							\bottomrule				     			},
				     },
every even row/.style={
					before row={\rowcolor[gray]{0.9}
							  }
				  },				     
  every last row/.style={after row=\hline},
  ]{./table/data/main-table.csv}

\label{table:mainvultable}
\end{table*}

This table demonstrates the number of vulnerabilities detected in nonOS packages i.e., application and its dependencies packaged in 55 Docker images. The column ``highest" (in table~\ref{table:mainvultable}) presents most vulnerabilities detected in a package, whereas, the column ``most vulnerable package" represents its corresponding package name. We can observe that in most cases one package is the main culprit in contributing to the total number of vulnerabilities detected.  For example., P1 has 128 vulnerabilities detected in 4 packages (or dependencies), in which, 120 different vulnerabilities are detected in MySQL package, which represents more than 90\% of the total detected vulnerability in P1. These vulnerabilities in MySQL package is because of Denial of Service (DoS) attack, which can cause the system to hang or crash.  

Our analysis also show that the most vulnerable Docker image is P39, which is affected by 1085 different vulnerabilities. Around 50\% of different vulnerabilities in P39 is again because of MySQL package.  We can observe from table~\ref{table:mainvultable} that attack surface on MySQL packages is high compared to other packages. For example, P29 has vulnerabilities detected in two packages with 262 different vulnerabilities, out of which, 261 vulnerabilities are because of MySQL. On further investigation, we find that most of these vulnerabilities can be fixed by updating to a new version of MySQL.

We also find there are four images such P21, P27,  P49 \& P52 in which vulnerabilities are being detected in one package. This can easily be fixed by updating the package.  However, the lack of updates seem to suggest that developers are not using newer versions of nonOS packages.

Our analysis also shows that existing Docker images on DockerHub suffer entirely from vulnerabilities in nonOS packages i.e., dependencies of applications (as shown by ``dependencies'' column  in table~\ref{table:mainvultable}). Furthermore, we can clearly see that no vulnerability has been detected in the application package of all the analyzed images as shown by column ``application'' in table~\ref{table:mainvultable}. This is because the approach used by the existing container scanning tools rely on metadata information (e.g., package name, version). The tool detects a vulnerable package by mapping each detected package and its version onto the list of  package information with reported vulnerabilities in a database. However, list of known vulnerabilities in the database are often incomplete, or missing which can cause the tool to miss certain vulnerabilities~\cite{Serena2018}. Therefore, in the next section, we use code inspection technique with SpotBugs to identify vulnerabilities in application package.

\subsection{Identifying application code vulnerabilities with code inspection}

\begin{figure}[ht]
	\centering
	\includegraphics[scale=0.45]{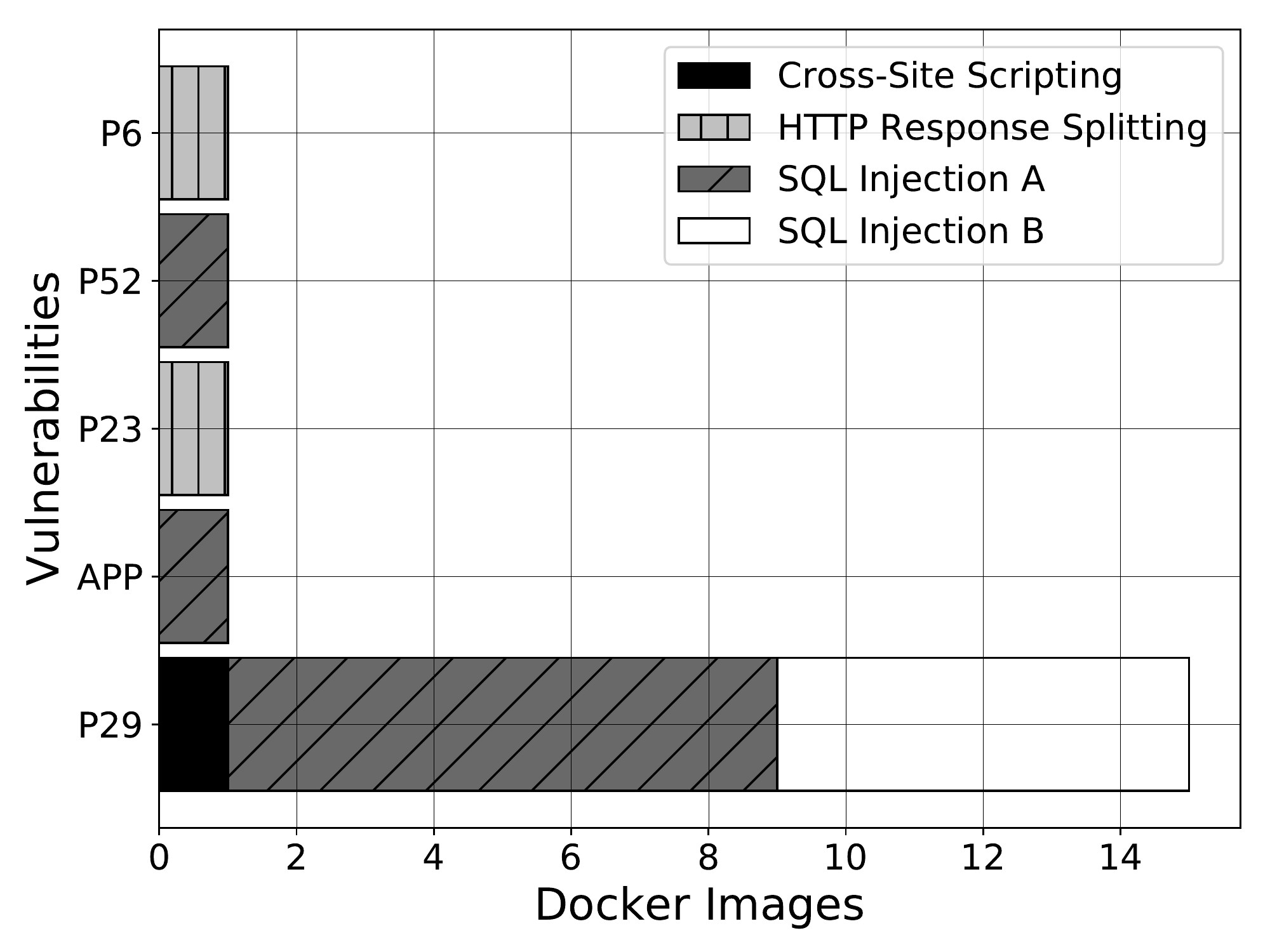}
	\caption{Vulnerabilities detected by code inspection on the application code of Docker Images. APP represents an application package vulnerability found in an image which was not detected  by Anchore.} 
	\label{fig:appvul}
\end{figure}
To detect vulnerabilities with code inspection, we use SpotBug's security plugin. We analyze the code of 59 Docker images. For the setup, we use the build system of Docker image's project on GitHub, which is based on either \textit{maven}, \textit{gradle} or \textit{ivy}. This eased the task of compiling and fetching the dependencies for  projects. Some projects failed the compilation process due to configuration issues, which was manually checked and fixed. We then used SpotBug's plugin to analyze the compiled Java classes. 

We find vulnerabilities in application code of 5 Docker image's project (as shown in figure~\ref{fig:appvul}). This figure shows a stacked bar chart with Docker images (i.e., its GitHub project) on x-axis, and number of detected vulnerabilities on the y-axis. There are in total 19 vulnerabilities detected in different application code of 5 projects. Out of these 19 vulnerabilities, 10 are because of SQL injection-A, 6  SQL injection-B, 2 HTTP Response Splitting, and 1 Cross-Site Scripting. Therefore, most of the vulnerabilities detected are due to SQL injection. For SQL related vulnerability, developers create a query by concatenating variables with a string. For example \code{"Select * FROM customers WHERE name = '`' + custName + '`'}. This makes the code vulnerable to SQL injection, because if an attacker gets hold of the system, she can concatenate malicious data to the query~\cite{thomas}.

We find P29 contains the highest number of application vulnerabilities. It has 15 vulnerabilities of 3 different kinds (i.e., cross-site scripting, SQL injection-A and SQL injection-B) in the application code. Furthermore, the most pervasive kind of vulnerability that has been detected is SQL injection. To differentiate between each detected vulnerability, we further categorize SQL injection into 2 categories such as `SQL injection-A' and `SQL injection-B'. SQL injection-A represents  vulnerability which occurs due to ``Nonconstant string passed to execute or addBatch method on an SQL statement". Similarly, category B represents ``a prepared statement is generated from a nonconstant String''.

Furthermore, we also found vulnerability in the application code of Docker image which was not detected by container scanning tool (i.e., Anchore). The name of the image is terracotta/terracotta-server-oss (shown as APP in figure~\ref{fig:appvul}), which is an official Terracotta Server repository having more than \numprint{50000} downloads on DockerHub. This is being missed by the container scanning tool, because its existing scanning approach relies on package information, which is present in the vulnerability database. With the obtained results, we can now answer our first research question:

\begin{itemize}
\item[] \textit{\textbf{RQ1: }Is existing container vulnerability scanning approach effective in finding nonOS package vulnerabilities??}
\end{itemize}

We find several key issues with the existing scanning approach. Firstly, few existing container scanning tools detect nonOS package vulnerability. In our analysis, only Anchore could analyze nonOS package because it uses vulnerability feed containing nonOS package information. We also find a high number of vulnerabilities (\numprint{12357}) detected in nonOS packages (in  747 dependencies). This makes an average of ~17 vulnerabilities per package, demonstrating that nonOS packages (i.e., dependencies) used in Docker image are prone to security attacks. Therefore, tools that are not detecting nonOS package vulnerabilities, like Clair and Microscanner, will not be effective to conduct a complete vulnerability detection.
 This  which are being missed by other container scanning tools such as Clair and Microscanner.  

Secondly, existing container scanning tools rely on metadata associated with libraries (e.g., package name, version) in a vulnerability database. This information is used to identify whether the analyzed package is vulnerable or not. However, if the metadata information is not available, it will fail to detect a vulnerability. It is  because of this reason that Anchore missed 19 vulnerabilities in the application package of 5 different Docker images. Therefore, existing scanning tools are missing out on vulnerabilities that are present in 8.5\% (5 out of 59) of images in our dataset. Hence, our analysis highlights a limitation in existing container scanning approach for detection coverage.

We now investigate accuracy and completeness of the tools by detecting OS package related vulnerabilities.

\subsection{Detecting OS vulnerabilities in Docker images}

\begin{figure}[ht]
	\centering
	\includegraphics[scale=0.55]{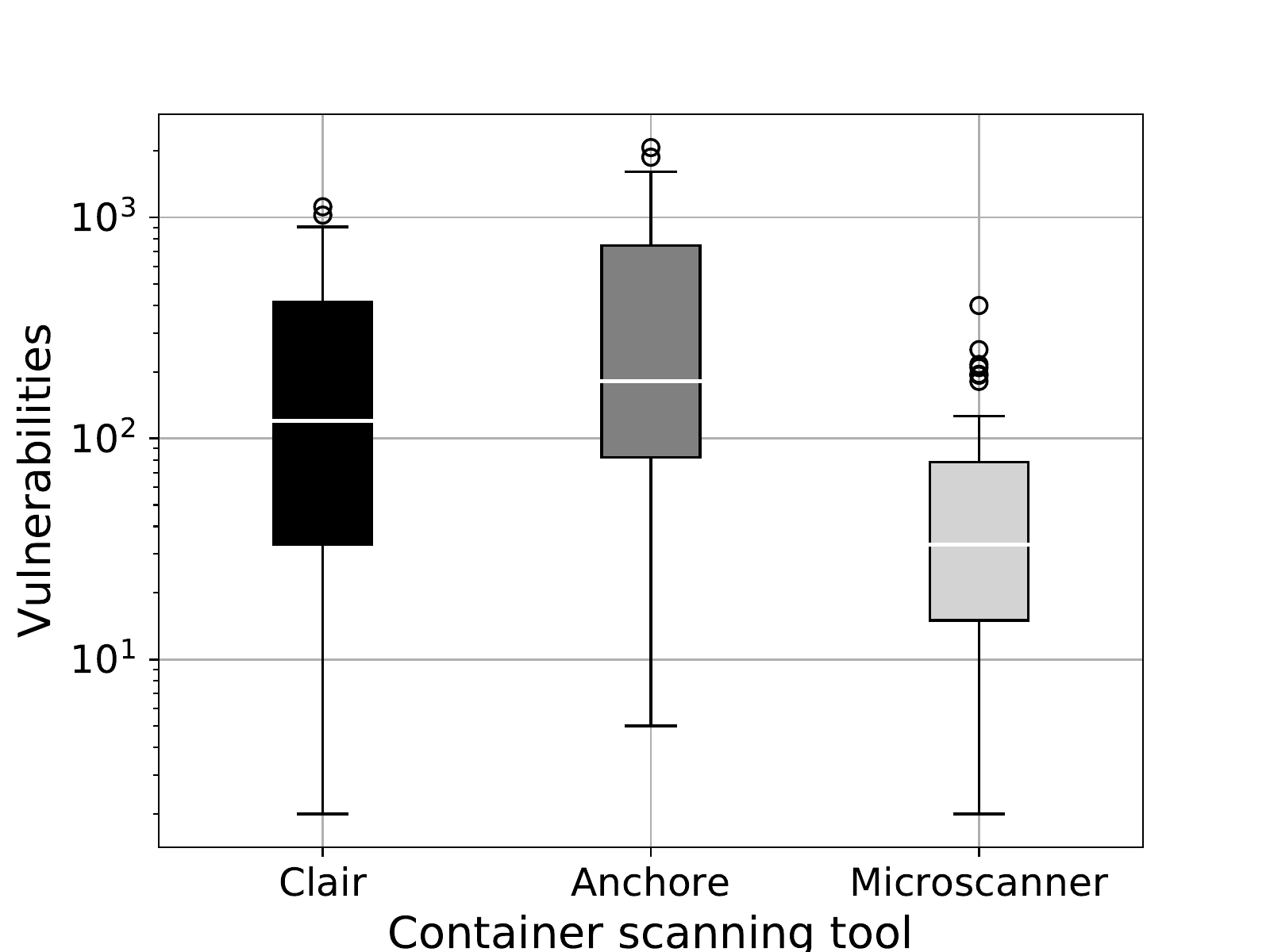}
	\caption{Comparison of vulnerability detection of container scanning tools.} 
	\label{fig:boxplot}
\end{figure}

To understand vulnerabilities related to OS packages, we analyze the same set of 59 Docker images using Anchore, Clair and Microscanner. Figure~\ref{fig:boxplot} shows the number of OS package vulnerabilities detected by all container scanning tools. X-axis represents the container scanning tools, whereas y-axis shows the number of detected vulnerabilities. We can observe that the mean vulnerabilities of the tools are 115 for Clair and 174 for Anchore. However, Microscanner has a very low mean value of 31. 

Furthermore, the lower whisker for Clair and Microscanner are same (i.e., 1 vulnerability detected). However, the upper whisker is different for all the tools, which means that there is very high dissimilarity among the detection of OS package vulnerabilities between the tools.

We investigated the reason for Anchore's higher vulnerability detection compared to Clair. We observe that anchore-engine tries to find OS package artifacts (apkg, rpm, dpkg)  during image analysis, which are then checked by upstream OS vendor security data sources to find potential vulnerability records. If the engine finds the package (e.g., curl) as vulnerable in data source, it will report all the libraries related to curl as vulnerable (e.g., libcurl3) using the same CVE identifier. Clair (on the other hand) reports only the main package (e.g., curl) as vulnerable and does not report the related libraries. Furthermore, the reason for lower values for Microscanner is that it misses many of the fixed vulnerabilities which are being detected by Clair and Anchore. This is because Microscanner database fetches vulnerabilities from different feeds compared to other tools. Microscanner does not provide much information about its vulnerability database, therefore, one possible reason could be that the database is not updated with package information.

\subsection{Finding detection miss for container scanning tools}
\label{miss}
In order to compute detection hit ratio, we find vulnerabilities detected and vulnerabilities which are missed by the tool. We find the number of detection by scanning the images, and filtering the detected vulnerabilities based on whether it was  fixed or not. For OS package vulnerability detection, each tool’s database has a field for vulnerability status (i.e., fixed or not). Therefore, we use this field to filter fixed vulnerabilities.

For finding detection miss by container scanning tools, we formally describe our procedure. Given the set of Docker images (i.e., 59 in this study), let $C_{i}$  (where $i = 1...n  \mid $ n=3 for this study) be the number of detected OS vulnerabilities by the tool. Each $C_{i}$  gets package information from its database. 

Elements of $v_{i} $ are in the form of \{image name, package name, package version, CVE identifier\}. $F_{v}$ $\mid$ $v_{f} \subseteq  v_{i}$ is a filtered set that contains only set of fixed vulnerabilities. We formalize our procedure in figure~\ref{fig:procedure}

\begin{figure}[ht]
\begin{algorithm}[H]
 \caption{Finding detection miss by container scanning tools. }
 \begin{algorithmic}[1]
 \renewcommand{\algorithmicrequire}{\textbf{Input:}}
 \renewcommand{\algorithmicensure}{\textbf{Output:}}
 \REQUIRE Set of fixed vulnerabilities.
 \ENSURE  Set of detection miss by container scanning tools.
 \\
  \FOR {$\forall v_{a} \in F_{v}$}
  \FOR{$\forall v_{a+1} \in F_{v}$}
  \IF {($v_{a} \ne v_{a+1}$)}
  \STATE $ V_{m}  \leftarrow (v_{a}  \rightouterjoin v_{a+1}) $
  \ENDIF
    \ENDFOR	  
  \ENDFOR
 \RETURN $V_{m}$
 \end{algorithmic}
 \end{algorithm}

 \caption{General procedure for collecting vulnerability information from different vulnerability database information}
  \label{fig:procedure}
\end{figure}

In figure~\ref{fig:procedure}, we demonstrate a general procedure for finding a set of detection miss for container scanning tools. Furthermore, our procedure compares 
vulnerability $v_{a}$ detected by a tool with vulnerabilities detected by all other tools such as $v_{a+1}$ to $v_{a+n}$ etc. This is handled by the two loops (in line1 and line2 of figure~\ref{fig:procedure}). However, we avoid same set of vulnerabilities i.e., elements in $v_{a}$ by itself (line 3).  Right outer join operation $\rightouterjoin$  is applied to determine whether elements present in $v_{a+1}$  are not in $v_{a}$. This allows to determine vulnerabilities which are not detected by the current container scanning tool.  We will now compute detection ratios for each tool.

\subsection{Comparing detection hit ratio of the tool}

We describe detection hit ratio as a measurement to understand a tool's effectiveness for detecting OS package vulnerabilities. 
For this measurement, we collect all confirmed vulnerabilities (i.e., vulnerability where there is an indication of an available  patch) that were reported by all three tools. From this set of vulnerability, we then identify how many vulnerabilities are detected by a tool, and how many are missed.  

\begin{table}[ht]
\caption{Detection Hit Ratio of Container scanning tools}
\centering
\begin{filecontents*}{./table/data/dhr.csv}
tool,detectionhits,detectionmiss,dhr,dhrp
Clair,7215,12798,0.36,36.05
Anchore,13149,6864,0.66,65.7
Microscan,2617,17396,0.13,13.08
\end{filecontents*}
\pgfplotstabletypeset[
  col sep=comma,
  columns/tool/.style={
  		string type,
  		column name=\textbf{Tools}, 
		column type={l}
  },
  columns/detectionhits/.style={
  	string type,
	column name=\textbf{\#Detection Hits}, 
	column type={c}
	},
  columns/detectionmiss/.style={
  	string type,
	column name=\textbf{\#Detection Miss}, 
	column type={c}
	},		
   columns/dhr/.style={
  	string type,
	column name=\textbf{DHR}, 
	column type={c}
	},				
  columns/dhrp/.style={
  	string type,
	column name=\textbf{DHR(\%)}, 
	column type={c}
	},				
 every head row/.style={
  					before row={		
								\toprule
							  },
					after row=\hline
				     },
  every last row/.style={after row=\bottomrule},
  ]{./table/data/dhr.csv}

\label{table:dhr}
\end{table}

Table~\ref{table:dhr} shows the ``Detection Hit Ratio" or DHR for each tool. The worst detection capability among the three tool is that of Microscanner. The DHR for Microscanner is very low i.e., only 13.08\%.  On the other hand, Anchore shows the best DHR among the three tools which is 65.7\%. However, we can still observe from the table that all three tools miss many vulnerabilities, 35\% in the best case for Anchore. Anchore misses \numprint{6864} vulnerabilities. These vulnerabilities are detected by other scanning scanning tools.

\begin{table*}[ht]
\caption{Summary of complete vulnerabilities detected in each section of the Docker Image}
\centering
\begin{filecontents*}{./table/data/completevul.csv}
tool,nonosimages,apppackages,libpackages,nonosvulnerbility,osimages,ospackages,osvulnerability,CompleteVulnerability
Clair,0,0,0,0,55,184,\numprint{7215},\numprint{7215}
Anchore,55,0,747,\numprint{12357},55,352,\numprint{13149},\numprint{25506}
Microscanner,0,0,0,0,44,91,2617,\numprint{2617}
\end{filecontents*}
\pgfplotstabletypeset[
  col sep=comma,
  columns/tool/.style={
  		string type,
  		column name=\textbf{Tool}, 
		column type={l}
  },
  columns/nonosimages/.style={
  	string type,
	column name=\textbf{Images}, 
	column type={r}
	},	
  columns/apppackages/.style={	
    	string type,
	column name=\textbf{Application},
    	column type={r}
  },
    columns/libpackages/.style={	
    	string type,
	column name=\textbf{Library},
    	column type={r}
  },
    columns/nonosvulnerbility/.style={	
    	string type,
	column name=\textbf{NonOSVulnerability},
    	column type={r}
  },
  columns/osimages/.style={	
    	string type,
	column name=\textbf{Images},
    	column type={r}
  },
    columns/ospackages/.style={	
    	string type,
	column name=\textbf{OSPackage},
    	column type={r}
  },
    columns/osvulnerability/.style={	
    	string type,
	column name=\textbf{OSVulnerability},
    	column type={r}
  },
      columns/CompleteVulnerability/.style={	
    	string type,
	column name=\textbf{CompleteVulnerability},
    	column type={r}
  },
 every head row/.style={
  					before row={		
								\toprule
							  },
					after row=\bottomrule				     },
  every last row/.style={after row=\bottomrule},
  ]{./table/data/completevul.csv}

\label{table:complete}
\end{table*}

Based on our findings, we can now answer our second research question:

\begin{itemize}
\item[] \textit{\textbf{RQ2: } What is the accuracy of vulnerability detection of existing container scanning tools? }
\end{itemize}

Based on the metric ``detection hit ratio", we find that Anchore seems to have a better detection hit ratio (~65.7\%) compared to the other two tools. This is because Anchore has a frequent vulnerability update mechanism. The anchore-engine periodically fetches newer version of the vulnerability data, and if the vulnerability information appears in the database during the scan, the engine will report the analyzed package having a vulnerability. Nonetheless, we see from our analysis that all the tools miss considerable number of vulnerabilities, which compromises the tool's accuracy for vulnerability detection.

In the subsequent section, we will investigate complete vulnerability landscape.

\subsection{Complete Vulnerability Landscape}

Evaluating the tools based on our two proposed metric provided sufficient information for assessing the quality of Docker images for Java applications hosted on DockerHub. We demonstrate complete vulnerability landscape i.e., the number of vulnerabilities detected in OS  and nonOS package of the container. Rui et al.~\cite{Rui2017} studied vulnerabilities of OS package of the container to understand the landscape. However,  a container packages both OS and nonOS packages, therefore, analyzing only OS package is not sufficient to assess vulnerability landscape of DockerHub. We compute complete vulnerability landscape of 59 Docker images by using the following formula:
\[V_{c}  = V_{app} + V_{lib} + V_{os} \]	
where
\begin{description}
\item[---]  $V_{app}$ is the number of vulnerabilities detected in application.
\item[---]  $V_{lib}$ is the number of vulnerabilities detected in libraries. 
\item[---]  $V_{os}$ is the number of vulnerabilities detected in OS packages. 
\end{description}

Table~\ref{table:complete}, summarizes the complete vulnerability detection for each tool. As we have discussed that the detection coverage of each tool is quite limited because vulnerabilities in the application package has been missed by all tools. 

Anchore has a higher complete vulnerability detection because it is the only tool in our analysis that is able to analyze nonOS package.  Anchore finds vulnerabilities in 55 images in both nonOS and OS vulnerability detection analysis. However, Anchore does not find vulnerabilities in the same 55 images. There are two images which are different in OS analysis, than in nonOS vulnerability analysis. P16 and P26 does not have any vulnerability in its OS packages. However, we find \code{krambox/knx2mqtt:latest} and \code{terracotta/terracotta-server-oss:latest} which contains OS vulnerability. Anchore did not detect nonOS package vulnerability in these two images. 


We can now address our last research question.
\begin{itemize}
\item[] \textit{\textbf{RQ3: }What is the complete vulnerability landscape i.e., vulnerabilities detected in both OS and nonOS packages of community images for Java-based container applications on DockerHub?}
\end{itemize}

When analyzing 59 Docker images, we find 19 vulnerabilities in application of 5 images. We only find $V_{lib}$ with Anchore, and detected \numprint{12357} in 747 library packages of 55 images. Clair and Anchore missed these vulnerabilities. Furthermore,  $V_{os}$ for Clair, Anchore and Microscanner are \numprint{7215}, \numprint{13149} and \numprint{2617} respectively. Thus, making complete vulnerability $V_{c}$  for Clair, Anchore, and Microscanner (i.e., \numprint{7215}, \numprint{25506} and \numprint{2617}) for the analysis of 59 Docker images. This shows a very high number of vulnerabilities exists in Docker images for Java applications hosted on DockerHub.

\section{Discussion}\label{discussion}

In this section, we discuss some key points about our metrics.

When analyzing 59 Docker images no vulnerabilities were reported in the application package. Based on the download information, we can see that the images are used by many users or developers (in the order of 700,000). A security exploit in any image could affect a large user base. Furthermore, our analysis demonstrated that existing container scanning tools will miss vulnerabilities in the application package. 19 vulnerabilities in application code of 5 images were detected by code inspection. We find vulnerability in one of the Docker project, which passed Anchore's nonOS package analysis. Therefore, dependency on existing vulnerability information by container scanning approach makes the detection coverage of existing container scanning tools limited. 

We also use Detection Hit Ratio (DHR) as a metric to understand the accuracy of the container scanning tools. We find that all  three tools did not demonstrate a good accuracy in detecting vulnerabilities. Microscanner performed poorly with only 13\% DHR. Clair and Anchore had higher DHR with $\approx$36\% and $\approx $65\% respectively. Nonetheless, in spite of Anchore's higher DHR, the tool is still missing around $\approx$34\% of vulnerabilities. Therefore, better vulnerability feed-sources are required, which should contain an updated vulnerability information of the packages.

From our analysis of detecting vulnerabilities in both OS and nonOS packages, we find that nonOS packages have on average $\approx$17 vulnerabilities (\numprint{12357} vulnerabilities in 747 packages). In case of OS packages, they have on average  ~37 vulnerabilities (13149 in 352). This shows that vulnerabilities detected in OS packages are  2.1$\times$ more than vulnerabilities detected in nonOS packages, thus making OS the more vulnerable part of the container.

\section{Threat to Validity}\label{threadvalid}

Since our study is analyzing constantly evolving Docker images, our evaluation is inherently time dependent. Performing the same analysis in another time (based on the updates to the images) may likely change our findings. We handle this problem by storing the exact version of the ``tag'' value along with the image name. This way, we try to achieve provenance of data so that the results can be reproduced. Furthermore, we acknowledge that our conclusions are valid only for applications written in Java, and may not be applicable to another programming language~\cite{Shahriar}. Therefore, future studies can expand this analysis to other languages.

Furthermore, our study focuses on reporting vulnerabilities in OS and nonOS packages in the container. We do not identify vulnerabilities in Docker (container) code. The reason is that vulnerabilities occurring in  Docker code are not detected by existing container scanning tools. Clair and Anchore only identify vulnerabilities in the packages. Furthermore, Docker code vulnerabilities are mostly not related to code issues (such as incorrect permissions or unprotected resources), making them hard to detect with static analysis tools~\cite{duarte}.

\section{Recommendations}~\label{recommendation}

Based on the experience from this study, we recommend few suggestions that can improve the vulnerability detection technique of container scanning tools.

\paragraph{Better detection approach} As our study shows that reliance on package vulnerability database can miss vulnerabilities in a container. This is because the existing approach detects vulnerabilities that have been reported or fixed. Therefore, we recommend the usage of a hybrid approach by using metadata to find reported vulnerabilities (as identified from a vulnerability database), and by running static code analysis to detect new types of vulnerabilities~\cite{KuhnRK}. This will allow the effective removal of security issues from OS and nonOS packages present  in the container.

\paragraph{Expanding detection coverage} Code inspection technique for vulnerability detection used in this study uses static code analysis to detect vulnerabilities. 
However, there are several disadvantages of static code vulnerability detection technique. First, it does not take into account the dynamic behavior of the application i.e., whether the vulnerable code will be covered~\cite{Serena2018} during the execution of the code. Second, static analyzer struggles with object-oriented languages (e.g., Java uses reflection to modify code during run-time)~\cite{Serena2018}. Therefore, we recommend the use of Dynamic Program Analysis (DPA) to detect vulnerabilities of a running program~\cite{amel}. This is because recent advancements in the instrumentation techniques have shown increased instrumentation coverage to detect potential bugs in Java standard class libraries at runtime~\cite{omar16}. This, therefore, can further expand vulnerability detection coverage.

\paragraph{Using real-world projects for evaluation} One important aspect of understanding the effectiveness of any bug detection tool is to evaluate the tool on a benchmark. Benchmarking of software security vulnerabilities allows to better understand software development practices and processes~\cite{Rotella}.  However, there is a lack of security benchmark for evaluating vulnerability detection tools. One of the available security benchmark is Open Web Application Security Project (OWASP) benchmark~\cite{Benchmar6}. It evaluates the accuracy of security detection tools. However, there are certain limitations of this benchmark, such as, it only supports static analysis testing tools. Furthermore, the benchmark is only limited to web-services. Therefore, we recommend that the research community can use large number of community based container images to understand the effectiveness of container scanning tools. Similar to previously proposed dataset for security analysis~\cite{8530718}, we provide a dataset consisting of vulnerability information of 59 popular Docker images at all levels of package granularity i.e., nonOS (application, and dependencies) and OS packages. This can be used by researchers and practitioners for evaluating vulnerability detection tools and techniques.

\section{Conclusion}\label{conclusion}

The use of containers present a security concern for which a number of container scanning tools have been developed. In this paper, we investigate the effectiveness of container scanning tools based on two metrics which represents coverage, and accuracy respectively. We investigate detection coverage of container scanning approach i.e., vulnerabilities detected in application and its dependencies packaged in the containers. Secondly, we look at the tool's effectiveness in terms of Detection Hit Ratio (DHR) to understand the accuracy in finding vulnerabilities. Finally, we studied the vulnerability landscape i.e., vulnerabilities detected in OS and nonOS package of real-world Docker images for Java applications hosted on DockerHub.  In assessing detection coverage, our findings indicate that vulnerabilities are missed by the existing vulnerability detection approach, because of its dependency on a vulnerability database. Secondly, we find that all the tools do not exhibit high DHR which means that they will fail to detect a high number of vulnerabilities. Finally, we find that they are more vulnerable OS packages when compared to nonOS packages.  Based on our experience from this study, we encourage the research community to focus on improving and developing better container vulnerability detection approach. 

\bibliographystyle{IEEEtran}
\bibliography{IEEEabrv,biblio}

\end{document}